\documentclass[
reprint,
superscriptaddress,
amsmath,amssymb,
aps,
pra,
]{revtex4-2}

\usepackage{graphicx}
\usepackage{epstopdf}
\usepackage{xcolor}

\usepackage{bm}
\usepackage{gensymb}
\usepackage{siunitx}

\usepackage{array}
\usepackage{booktabs}
\usepackage{dcolumn}
\usepackage{tabularx}
\usepackage{multirow}

\usepackage{float}
\usepackage{subfig}

\usepackage{hyperref}
\hypersetup{
    colorlinks=true,
    linkcolor=blue,
    anchorcolor=blue,
    citecolor=blue,
    urlcolor=blue
}

\usepackage{orcidlink}
\usepackage{tikz}
\usepackage{mwe}

\usepackage{textcomp}
\usepackage{verbatim}
\usepackage{url}

\newcolumntype{Y}{>{\centering\arraybackslash}X}
\newcolumntype{x}{D{.}{.}{6.6}}
\newcolumntype{y}{D{.}{.}{4.5}}
\newcolumntype{z}{D{.}{.}{5.7}}
\newcolumntype{f}{D{.}{.}{7.9}}
\newcolumntype{e}{D{.}{.}{5.6}}

\usepackage{upgreek}

\begin{document}
\preprint{Online PLASEN}
\title{Commissioning and Full Realization of the PLASEN System at BRIF}

\author{W.~C.~Mei\orcidlink{0009-0009-7635-6998}}
\altaffiliation{These authors contributed equally to this work.}
\affiliation{School of Physics and State Key Laboratory of Nuclear Physics and Technology, Peking University, Beijing 100871, China}

\author{H.~R.~Hu\orcidlink{0009-0006-7760-3338}}
\altaffiliation{These authors contributed equally to this work.}
\affiliation{School of Physics and State Key Laboratory of Nuclear Physics and Technology, Peking University, Beijing 100871, China}

\author{Y.~F.~Guo\orcidlink{0009-0009-1817-7959}}
\altaffiliation{These authors contributed equally to this work.}
\affiliation{School of Physics and State Key Laboratory of Nuclear Physics and Technology, Peking University, Beijing 100871, China}

\author{Z.~Yan\orcidlink{0009-0007-8129-4600}}
\altaffiliation{These authors contributed equally to this work.}
\affiliation{School of Physics and State Key Laboratory of Nuclear Physics and Technology, Peking University, Beijing 100871, China}

\author{X.~F.~Yang\orcidlink{0000-0002-1633-4000}}
\email{xiaofei.yang@pku.edu.cn}
\affiliation{School of Physics and State Key Laboratory of Nuclear Physics and Technology, Peking University, Beijing 100871, China}

\author{S.~J.~Chen\orcidlink{0009-0005-1235-2411}}
\affiliation{School of Physics and State Key Laboratory of Nuclear Physics and Technology, Peking University, Beijing 100871, China}

\author{D.~Y.~Chen\orcidlink{0000-0001-6264-060X}}
\affiliation{School of Physics and State Key Laboratory of Nuclear Physics and Technology, Peking University, Beijing 100871, China}

\author{Y.~P.~Lin}
\affiliation{School of Physics and State Key Laboratory of Nuclear Physics and Technology, Peking University, Beijing 100871, China}
\author{Y.~S.~Liu}
\affiliation{School of Physics and State Key Laboratory of Nuclear Physics and Technology, Peking University, Beijing 100871, China}
\author{C.~Zhang}
\affiliation{School of Physics and State Key Laboratory of Nuclear Physics and Technology, Peking University, Beijing 100871, China}
\author{Y.~P.~Jing}
\affiliation{School of Physics and State Key Laboratory of Nuclear Physics and Technology, Peking University, Beijing 100871, China}
\author{T.~X.~Gao\orcidlink{0009-0008-1025-6495}}
\affiliation{School of Physics and State Key Laboratory of Nuclear Physics and Technology, Peking University, Beijing 100871, China}

\author{X.~Shen}
\affiliation{School of Physics and State Key Laboratory of Nuclear Physics and Technology, Peking University, Beijing 100871, China}
\author{Y.~Y.~Jia}
\affiliation{School of Physics and State Key Laboratory of Nuclear Physics and Technology, Peking University, Beijing 100871, China}
\author{Y.~T.~Lin}
\affiliation{School of Physics and State Key Laboratory of Nuclear Physics and Technology, Peking University, Beijing 100871, China}
\author{H.~X.~Zhang}
\affiliation{School of Physics and State Key Laboratory of Nuclear Physics and Technology, Peking University, Beijing 100871, China}
\author{S.~W.~Bai\orcidlink{0000-0002-6087-9788}}
\affiliation{State Key Laboratory of Nuclear Physics and Technology, Institute of Quantum Matter, South China Normal University, Guangzhou 510006, China}

\author{B.~Tang}
\email{tangb364@ciae.ac.cn}
\affiliation{China Institute of Atomic Energy, P.O. Box 275 (10), Beijing 102413, China}

\author{X.~Ma}
\affiliation{China Institute of Atomic Energy, P.O. Box 275 (10), Beijing 102413, China}
\author{G.~F.~Song}
\affiliation{China Institute of Atomic Energy, P.O. Box 275 (10), Beijing 102413, China}
\author{S.~Ye\orcidlink{0009-0001-0101-2631}}
\affiliation{China Institute of Atomic Energy, P.O. Box 275 (10), Beijing 102413, China}
\author{M.~Y.~Lu}
\affiliation{China Institute of Atomic Energy, P.O. Box 275 (10), Beijing 102413, China}
\author{J.~Y.~Dong\orcidlink{0009-0007-6483-7335}}
\affiliation{China Institute of Atomic Energy, P.O. Box 275 (10), Beijing 102413, China}
\author{B.~K.~Dong}
\affiliation{China Institute of Atomic Energy, P.O. Box 275 (10), Beijing 102413, China}

\author{J.~H.~Lv}
\affiliation{Institute of Modern Physics, Chinese Academy of Sciences, Lanzhou 730000, China}
\author{S.~Y.~Dong}
\affiliation{School of Nuclear Science and Technology and Frontiers Science Center for Rare Isotopes, Lanzhou University, Lanzhou 730000, China}

\author{F.~C.~Liu}
\affiliation{Key Laboratory of Beam Technology of Ministry of Education, School of Physics and Astronomy, Beijing Normal University, Beijing 100875, China}
\author{Z.~Hu}
\affiliation{Key Laboratory of Beam Technology of Ministry of Education, School of Physics and Astronomy, Beijing Normal University, Beijing 100875, China}
\author{X.~Liu}
\affiliation{Key Laboratory of Beam Technology of Ministry of Education, School of Physics and Astronomy, Beijing Normal University, Beijing 100875, China}
\author{S.~T.~Zhu}
\affiliation{Key Laboratory of Beam Technology of Ministry of Education, School of Physics and Astronomy, Beijing Normal University, Beijing 100875, China}
\author{Y.~L.~Yi}
\affiliation{Key Laboratory of Beam Technology of Ministry of Education, School of Physics and Astronomy, Beijing Normal University, Beijing 100875, China}

\author{C.~Y.~He}
\affiliation{China Institute of Atomic Energy, P.O. Box 275 (10), Beijing 102413, China}

\author{A.~Takamine\orcidlink{0000-0001-5528-7940}}
\affiliation{RIKEN Nishina Center for Accelerator-Based Science, Wako, Saitama 351-0198, Japan}
\affiliation{Department of Physics, Kyushu University, Motooka, Nishi-Ku, Fukuoka, Fukuoka 819-0395, Japan}

\author{B.~Q.~Cui}
\affiliation{China Institute of Atomic Energy, P.O. Box 275 (10), Beijing 102413, China}
\author{J.~Yang}
\affiliation{Institute of Modern Physics, Chinese Academy of Sciences, Lanzhou 730000, China}
\author{Z.~Y.~Liu}
\affiliation{School of Nuclear Science and Technology and Frontiers Science Center for Rare Isotopes, Lanzhou University, Lanzhou 730000, China}
\author{J.~Su}
\affiliation{Key Laboratory of Beam Technology of Ministry of Education, School of Physics and Astronomy, Beijing Normal University, Beijing 100875, China}
\author{H.~N.~Liu}
\affiliation{Key Laboratory of Beam Technology of Ministry of Education, School of Physics and Astronomy, Beijing Normal University, Beijing 100875, China}

\author{Y.~L.~Ye\orcidlink{0000-0001-8938-9152}}
\affiliation{School of Physics and State Key Laboratory of Nuclear Physics and Technology, Peking University, Beijing 100871, China}

\author{B.~Guo}
\affiliation{China Institute of Atomic Energy, P.O. Box 275 (10), Beijing 102413, China}

\begin{abstract}
A PLASEN~(Precision LAser Spectroscopy for Exotic Nuclei) system, consisting of a compact radio-frequency quadrupole cooler–buncher~(RFQ-cb) and a collinear resonance ionization spectroscopy setup, has now been fully commissioned with radioactive ion beams at the Beijing Radioactive Ion-beam Facility ~(BRIF). Using both stable and radioactive Rb ion beams from BRIF, we demonstrated that the large beam energy spread observed at BRIF has been successfully handled by employing the RFQ-cb, enabling the delivery of high-quality bunched radioactive ion beams for collinear resonance ionization spectroscopy experiments. Under these conditions, we performed laser spectroscopy of exotic nuclei, achieving high resolution($\sim$100 MHz spectral linewidth) and high sensitivity~($\sim$1:200 efficiency). This fully operational PLASEN system will serve as a state-of-the-art experimental platform at BRIF for research in multiple fields such as nuclear, atomic and molecular physics.
\end{abstract}

\maketitle

\section{Introduction}\label{sec1}
\quad As a quantum many-body system governed by multiple interactions, the atomic nucleus exhibits a diversity of intricate structures, providing a wide variety of physical pictures that are complementary and mutually exclusive ~\cite{PhysofExoticNuclei}. Over the past few decades, numerous radioactive ion-beam (RIB) facilities have flourished worldwide, enabling artificial production and study of short-lived unstable nuclides. These advances have greatly extended the boundaries of the nuclear chart and deepened our understanding of nuclear structure and interactions. For unstable nuclei far from the valley of $\beta$-stability, remarkable exotic structural phenomena have been observed~\cite{PhysofExoticNuclei}, including the modification of tranditional nuclear magic numbers due to shell evolution~\cite{ShellEvolution,NewMagicNumber}, the emergence of the island of inversion ~\cite{IOIPRC1975,IOIPRC1990,IOIPhysics,IOIPRL2023,IOIPRL2023Zn}, shape coexistence~\cite{ShapeCoexistenceRMP2011,ShapeCoexistencePRL2016,ShapeCoexistencePRL2016Ni}, as well as the formation of nuclear halos~\cite{HaloEPL1987,HaloPRL1985,HaloRMP2004,HaloPS2013} and clusters~\cite{ClusterPR2006,ClusterNST2018,ClusterPRL2020} in weakly-bound nuclei near the dripline. These discoveries have drawn widespread attention and have become a major frontier in contemporary low-energy nuclear physics research.\par

\quad In particular, measurements of nuclear properties, such as mass, spin, electromagnetic moments and charge radius, have long provided essential insights into the exotic structure of unstable nuclei~\cite{LaserSpectroscopyforExoticNuclei}. Moreover, such measurements can not only deepen our understanding of nuclear forces and the many-body correlations among nucleons, but also stimulate the development of \textit{ab initio} theoretical calculations in nuclear physics~\cite{abinitio1,abinitio2}. At the interface between nuclei and atomic systems, nuclear electromagnetic properties induce perturbations on the motion of extranuclear electrons, giving rise to subtle yet measurable frequency shifts and hyperfine splittings, as small as one in a million of the total transition frequency, in atomic and molecular spectra. High-resolution laser spectroscopy of these hyperfine structures and isotope shifts in atoms or molecules enables the extraction of fundamental nuclear properties, including spins, electromagnetic moments, and root-mean-square charge radii in a nuclear-model-independent manner~\cite{LaserSpectroscopyforExoticNuclei}, providing a unique tool for exploring exotic structures and underlying effective interactions that emerge or are enhanced in unstable nuclei.\par

\quad To study unstable species at RIB facilities, a variety of advanced laser spectroscopy techniques have been developed and applied across the nuclear chart~\cite{LaserSpectroscopyforExoticNuclei}. These include in-source resonance ionization spectroscopy~\cite{Insource1,Insource2}, collinear laser spectroscopy (CLS) with laser-induced fluorescence (LIF) detection~\cite{COLLAPS,BECOLA,REBELandSTRIPE}, collinear resonance ionization spectroscopy (CRIS)~\cite{CRIS,hm9t-4pgp}, and several trap-based approaches employing atoms or ions~\cite{REBELandSTRIPE,HeliumatArgonne,OROCHI,MIRACLS}. Each technique offers distinct advantages in precision, resolution or detection sensitivity, making them suitable for different scientific goals. Among them, the collinear resonance ionization spectroscopy, which combines the high-resolution of collinear laser-beam interactions with the high-sensitivity of resonance-ionization detection, has proven to be a state-of-the-art technique for studying nuclear properties across various regions of the nuclear chart~\cite{LaserSpectroscopyforExoticNuclei,CRIS,hm9t-4pgp}.\par

\quad Aiming to perform laser spectroscopy of exotic nuclei at currently operational and newly-constructed RIB facilities~\cite{NuclPhysatBRIF, HIAF} in China, we have been developing high-resolution and high-sensitivity CLS techniques based on LIF and RIS detection. First-stage CLS setups employing LIF detection were constructed at Peking University (PKU)~\cite{offlineCLS} and at the Beijing Radioactive Ion-beam Facility (BRIF) of the China Institute of Atomic Energy (CIAE)~\cite{onlineCLS}, where offline and online commissioning were carried out respectively. In the online experiment at BRIF, the hyperfine structure (HFS) spectrum of radioactive $\rm{^{38}K}$ ($T_{1/2}=7.651(19) ~\mathrm{min}$) was successfully measured, although with limited efficiency and spectral resolution due to the continuous ion beam from BRIF and its relatively large energy spread~\cite{onlineCLS}. To overcome these limitations, we subsequently upgraded the CLS setup to incorporate RIS measurement~\cite{CLS-RIS-Zhang2023} and developed a radio-frequency quadrupole cooler-buncher (RFQ-cb) which enables the production of bunched beams with excellent ion-beam quality~\cite{offlineRFQ}. Building on this foundation, we completed a fully functional offline PLASEN (Precision LAser Spectroscopy for Exotic Nuclei) system, whose capabilities were validated through the successful measurement of the HFS spectra of stable $\rm{^{85,87}Rb}$ isotopes, achieving a high efficiency of 1:200 and a spectral resolution of about 100~MHz~\cite{offlinePLASEN}.\par

\quad Nevertheless, whether the RFQ-cb can effectively reduce the large energy spread of radioactive beam at BRIF, and whether the PLASEN system can fulfill its goal of performing high-resolution and high-sensitivity laser spectroscopy of unstable nuclei, remain to be confirmed at a RIB facility. Recently, we installed the entire PLASEN system, comprising the RFQ-cb and the upgraded CLS setup, and carried out the first successful collinear resonance ionization spectroscopy experiment on unstable nuclei at BRIF, thereby demonstrating the full functionality of the system.

\quad In this work, we will introduce the online PLASEN system at BRIF and report the first fully-commissioning experiment performed with neutron-rich Rb isotopes. A detailed description of the experimental system's setup and performance is provided in the following sections.\par

\section{Online PLASEN system at BRIF}\label{sec2}

\begin{figure*}[htbp]
    \centering
    \includegraphics[width=0.98\textwidth]{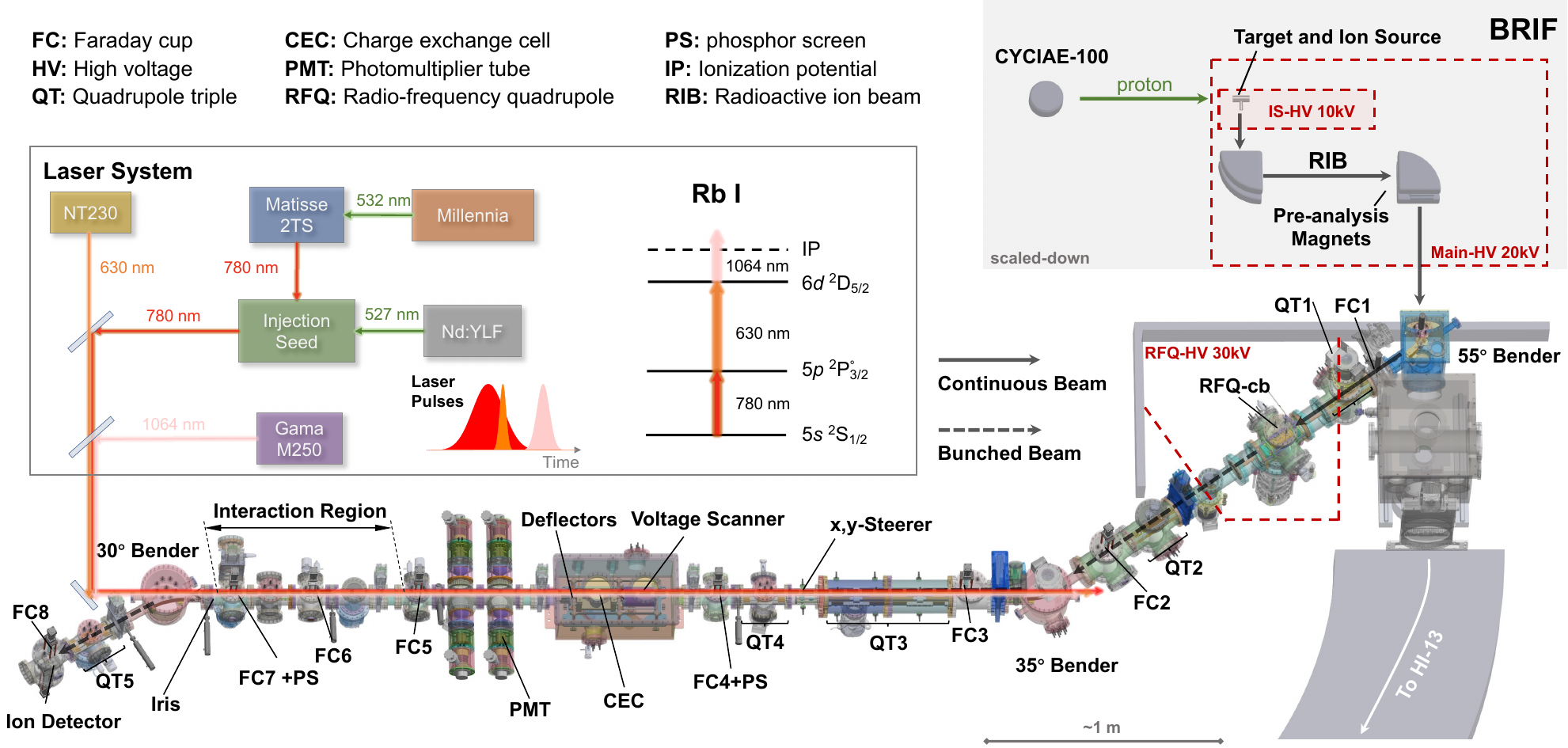}
    \captionsetup{justification=raggedright}
    \caption{Schematic layout of the online PLASEN system at BRIF. A 100-MeV proton beam bombards a solid target at BRIF to produce radioactive isotopes, which are ionized, accelerated and mass-separated. This continuous ion beam from BRIF is then cooled and bunched by the RFQ-cb. In the CLS beamline, ion bunches are neutralized in the charge-exchange cell, and then resonantly re-ionized by three step laser beams in the interaction region. The re-ionized ion bunches are detected to obtain the hyperfine-structure spectrum. The BRIF shown in the top-right inset is schematically scaled down and not to scale. See text for details.}
    \label{fig1}
\end{figure*}

\quad Figure~\ref{fig1} presents the overall layout of the online PLASEN system at BRIF, including the RFQ-cb, the CLS setup, and the associated laser system. A large number of radioactive isotopes can be produced when a 100-MeV proton beam impinges on a solid target at BRIF. The nuclear reaction products are then ionized in an ion source, extracted and accelerated to energies of about 30 keV. The isotope of interest can be  selected by the isotope separator and delivered to experimental terminals. Prior to delivery to the CLS setup, the continuous ion beam from BRIF is cooled and bunched by the RFQ-cb. Upon reaching the collinear beamline, the bunched ion beam will be guided to the charge-exchange cell (CEC), where it is neutralized by passing through dense alkali-metal (Na or K) vapor. Within the interaction region, the resulting neutral atom bunches are time-synchronized and spatially overlapped with multiple laser beams. After photoionization, the re-ionized ion bunches are directed toward the ion detector. The count rates of the ions are recorded by the data-acquisition (DAQ) system as a function of the laser frequency, yielding the HFS spectrum of the studied isotope.

\subsection{BRIF}\label{sec2.1}
\quad BRIF is an ISOL (Isotope Separator On-Line) -type facility capable of providing a wide variety of low-energy radioactive ion beams~\cite{NuclPhysatBRIF}. To produce radioactive nuclides, a 100-MeV high-intensity proton cyclotron (CYCIAE-100) is used to irradiate thick solid targets composed of heavy actinide compounds, thereby inducing nuclear reactions within the target.\par

\quad By heating the ISOL targets and directly-connected ion source to typically $2000~\mathrm{^{\circ} C}$, radioactive nuclides with low melting points are released from the targets and subsequently ionized inside the ion source. The ion source at BRIF can operate in dual modes: as a surface ion source or as a forced electron beam induced arc discharge (FEBIAD) ion source. Surface ionization is suitable for elements with relatively low ionization potentials (IP), such as alkali and alkaline-earth metals, whereas FEBIAD ion source is more effective for elements with comparatively higher ionization potentials.

\quad The ions continuously extracted from the ion source are initially accelerated to 10 keV by the first high-voltage (HV) stage. The ion beam is then transported into a set of high-resolution dipole magnets with a mass-resolving power of about 2000 in the pre-analysis section for isotope separation. As the pre-analysis section is floated at 20 kV on the main HV platform, the mass-selected ion beam is further accelerated to reach a total energy of 30 keV, as shown by the top-right inset of Fig.~\ref{fig1}. The ion beam then passes through a 55$^{\circ}$ electrostatic bender to enter the PLASEN experimental setup.\par

\subsection{RFQ cooler buncher}\label{sec2.2}
\quad The first major component of the PLASEN system is an RFQ-cb, a linear Paul trap. It consists of radio-frequency quadrupole electrodes, twelve DC ring electrodes, as well as cone and cap electrodes for ion injection and extraction. The entire RFQ-cb assembly is floated on a 30-kV HV platform powered by a high-precision DC power supply (Heinzinger PNChp 40000-15pos) with the stability typically better than 0.001\% over 8 hours. A high-precision voltage divider (Ohm-Labs, KV-30A), connected to the RFQ-cb chamber and read out by a 7.5-digit high-precision digital multimeter (Keysight, 34470A), continuously monitors and records the accurate beam energy in real time. 

\quad The radioactive ion beam from BRIF, with an energy of about 30 keV, is decelerated and injected into the RFQ, where it collides with helium of 99.999$\%$ purity, dissipating kinetic energy and thereby being cooled by the buffer gas. Two RF signals with a 90$^\circ$ phase difference are applied to the quadrupole electrodes, with identical voltages on opposite electrodes and opposite polarities on adjacent ones. According to the stability conditions of the Mathieu equation, an appropriate combination of RF voltage and frequency provides a transverse confining potential that focuses the beam along the center axis of the quadrupole electrodes and thus reduces the emittance of the extracted ion beam. Meanwhile, the injection and extraction caps, together with the 12 DC-electrodes, generate a longitudinal confining potential that restricts the ion motion within the trap. In particular, the extraction cap electrode is connected to a HV switch (Behlke, GHTS30) that enables rapid polarity reversal of the applied voltage, controlled by a 100-Hz TTL signal generated by a pulse generator (Quantum Composer, 9520). This configuration allows control of ion accumulation and release, thus producing a bunched beam. Further details of the RFQ-cb, including the power supplies, buffer gas flow control and optoelectronic communication with the devices on the HV platform, can be found in our previous work ~\cite{offlineRFQ}.
 
\quad The temporal width of the generated pulsed beam and the corresponding transmission efficiency of the RFQ-cb strongly depends on the electrode setting. In our earlier offline work, a pulsed beam with a temporal width of about 2 $\mathrm{\upmu s}$ (FWHM: Full Width at Half Maximum) and a transmission efficiency of approximately 60$\%$ were achieved~\cite{offlineRFQ}. However, in the present online experiment, although the same temporal width of 2 $\mathrm{\upmu s}$ can be easily reached, the transmission efficiency of the online RFQ-cb could not be directly determined since the first Faraday cup~(FC1) upstream of the RFQ-cb is only a simplified copper plate due to the limited installation space at BRIF. Nevertheless, a total transmission efficiency exceeding 30\% was obtained from the second dipole magnet of BRIF to the FC downstream of the RFQ-cb, which includes both the transport efficiency along the BRIF beamline and the transmission through the RFQ-cb, as shown in Fig.~\ref{fig1}. It is worth noting that the ion beam spot delivered from BRIF to the 55$^{\circ}$ bender was relatively large, as evidenced by the beam current measured at the 30-mm-diameter annular  segmented plate centered on the beam axis. The large beam size consequently limited the transmission efficiency through the 5-mm aperture in front of the RFQ-cb~\cite{offlineRFQ}. 

\subsection{CLS setup}\label{sec2.3}
\quad After release from the RFQ-cb, the bunched ion beam is reaccelerated to 30 keV and guided through a 35$^{^\circ}$ electrostatic bender into the CLS beamline, as shown in Figure~\ref{fig1}. It then passes through a series of ion optics, including x–y steerers, quadrupole triplet lenses (QT). These ion optics electrodes are powered by $\pm$6 kV DC power supplies (iseg, EHS 80 60n/p and EHS F0 60n/p) housed in a HV crate (iseg, ECH54A), allowing high-precision voltage control. Multiple beam-diagnostic devices, such as FC, phosphor screens (PS), and iris diaphragms are installed along the collinear beamline. Most FCs are equipped with suppression electrodes biased at -50~V to reduce secondary-electron emission. The ion-beam intensity is measured and monitored using a picoammeter (Keithley, 6485), and all the ion optics and beam-diagnostic devices for beam-tuning can be remotely controlled by dedicated software~\cite{PKUDAQ,Yanzhou2025}~(see Sec.~\ref{sec2.5} for details).\par

\quad Upon reaching the CEC, the ions undergo charge-exchange reactions with dense sodium vapor, resulting in neutralization~\cite{onlineCLS}. The residual non-neutralized ions are subsequently removed from the beamline by downstream deflectors. Upstream of the CEC, a voltage-scanning electrode is electrically connected to the cell. By applying tunable voltages (up to $\pm$2000~V) to this electrode via a voltage amplifier (Trek, 623B DC), the beam energy can be continuously varied, enabling Doppler tuning as an alternative spectroscopic method to laser frequency-scanning. To obtain HFS spectra using Doppler tuning approach, the actual voltages applied to this scanning electrode must be recorded in real time with high precision and stability. This is accomplished by using a precision voltage divider (Ohm-Labs, KV-10A) and a digital multimeter (Keysight, 34470A).\par

\quad The neutralized atomic bunches emerging from the CEC then enter the meter-long interaction region through a series of apertures, where they encounter the multi-step laser pulses (Sec.~\ref{sec2.4}) under precise timing synchronization and spatial overlap. The interaction region is maintained under ultra-high vacuum (typically on the order of $10^{-9}\sim~10^{-10}$ mbar), achieved using a combination of vacuum-pump sets, non-evaporable getter pump, and differential pumping system~\cite{CLS-RIS-Zhang2023}. This is to minimize ion background from collisional ionization of neutral atoms. After being stepwise resonantly excited and ionized by the lasers, the re-ionized ions are guided toward the ion detector (ETP Ion Detect, MagneTOF Plus 149403) by a 30$^{\circ}$ electrostatic bender and a QT. The ion counts are recorded by the ion detector via the DAQ  system (Sec.~\ref{sec2.5}) as functions of arrival time and tuning voltage (or laser frequency) , yielding the time-of-flight (TOF) spectra and HFS spectra of the isotope of interest, respectively.\par

\quad It is worth noting that, in addition to the above-mentioned RIS measurements, the PLASEN system is also designed to perform CLS experiments using LIF detection. In such experiments, the neutral atomic beam is resonantly excited by a narrow-band single continuous-wave (CW) laser. The subsequent emitted fluorescence photons are detected by the photon detection system and recorded by the DAQ system as a function of the tuning voltage or laser frequency to obtain the HFS spectrum of the isotope of interest. Further details about the structure and the operation of this photon detection system are given in Refs.~\cite{offlineCLS, onlineCLS}.\par

\subsection{Control and DAQ system}\label{sec2.5}

\quad The ion-beam transport throughout the entire PLASEN system is controlled via a fully remote-operated control and DAQ system based on Experimental Physics and Industrial Control System (EPICS) framework and Python, which can also be automated using a machine-learning-assisted (ML-assisted) beam-tuning program~\cite{Yanzhou2025}. The DC HV required for the ion optics is mainly supplied by HV modules installed in two iseg crates with Ethernet/CAN interfaces. An ECH 54A crate equipped with a CC 24 MMS controller serves as the main control unit. It is connected to another ECH 224 crate located inside the 30-kV HV platform for the RFQ-cb through optical-electrical communication using the CAN interface, ensuring stable and safe data transmission between the two crates. All the above-mentioned hardware devices are operated by corresponding computer programs via serial or Ethernet interfaces whenever possible, enabling fully remote operation during experiments.

\quad Data acquisition and processing from the photomultiplier tube (PMT) photon detector or MagneTOF ion detector are handled by the DAQ system, which consists of hardware electronic modules for data preprocessing and Python-based software for data recording, storage and real-time visualization~\cite{PKUDAQ}. For instance, to ensure sufficient signal gain, the MagneTOF ion detector is biased at –2450~V, powered by a NIM-standard HV module (CAEN, N1470). The raw signal from the detector is processed through electronics modules that include a fast amplifier (ORTEC, FTA 820), a constant-fraction discriminator (CAEN, N605) and a level translator (Phillips, 726) before being fed into a time-to-digital converter (TDC) (cronologic, Timetagger4-2G-PCIe).

\quad The experimental timing scheme is organized as follows. The TTL signal controlling the extraction-cap electrode of the RFQ-cb, which releases the ion bunches, serves as the master trigger. This signal synchronizes the triggers of the three pulsed lasers and the gate signal of the TDC. By coordinating the timing in this way, the ion bunches and laser pulses overlap temporally in the interaction region, while the TDC gate corresponds to the TOF of the ions from their release at the RFQ-cb to their arrival at the detector. The DAQ software records the bunch number and TOF of each ion event, together with the scanning voltage and the applied laser frequency, including timestamps for subsequent synchronization analysis.

\quad The basic logic and operating principles of the control and DAQ system are described in detail in our previous work~\cite{PKUDAQ}. In the present experiment, the software was largely updated and optimized by incorporating more hardware devices, communication via EPICS as well as a ML-assisted beam-tuning program, which will be detailed in an upcoming work~\cite{Yanzhou2025}.\par

\subsection{Laser system}\label{sec2.4}
\quad As described above, measurement of HFS spectrum of the isotope of interest using RIS method requires multi-step pulsed laser beams, and consequently a complex laser system with precise frequency, power and timing control. In the present commissioning work, resonance ionization of Rb isotopes was performed, as detailed in Sec.~\ref{sec3.2}. Therefore, the three-step laser excitation and ionization scheme of Rb, shown in the Fig.~\ref{fig1}, is used here as an example to briefly introduce the laser system. Further technical details can be found in Ref.~\cite{offlinePLASEN}.\par

\quad In brief, the first  excitation step of Rb atoms, corresponding to the $5s~^2\mathrm{S}_{1/2}\rightarrow 5p~^2\mathrm{P}_{3/2}^{\circ}$ $\mathrm{D_2}$ transition, is driven by an injection-locked titanium-sapphire (Ti:Sa) laser operating at 100 Hz repetition rate with a pulse width of $\sim$50~ns and a linewidth of about 20~MHz. This laser is seeded by a narrow-band CW Ti:Sa laser and pumped by a 527-nm pulsed Nd:YLF laser. The laser wavelength is monitored throughout the measurement by a high-precision wavemeter (HighFinesse, WS-8), which is calibrated by a narrow-band CW diode laser locked to a specific hyperfine transition of Rb in a temperature-controlled vapor cell. The second-step $5p~^2\mathrm{P}_{3/2}^{\circ}\rightarrow 6d~^2\mathrm{D}_{5/2}$ transition is excited by a 100-Hz pulsed optical parametric oscillator~(OPO) laser (EKSPLA, NT230), whose output wavelength is factory-calibrated by a spectrometer. A high-power 1064-nm Nd:YAG laser (Beamtech, Gama-M250) is then used to excite the Rb atoms from $6d~^2\mathrm{D}_{5/2}$ to energies above the ionization potential.

\quad During the experiment, the frequency of the first-step laser was either locked to a fixed value corresponding approximately to Doppler-shifted $\mathrm{D_2}$ transition frequency of one Rb isotope for Doppler tuning (voltage scan) or scanned across the Rb's HFS for laser-frequency scan~\cite{offlinePLASEN}. The wavelength of the second-step laser was optimized to maximize the resonance-ionization ion counts. The pulse timing of the three-step lasers was monitored using a set of the photodiodes and an oscilloscope. Timing synchronization of all three pulsed lasers was achieved by operating them in external-trigger mode, with the timing setting adjusted and controlled via TTL signals from a pulse generator. Further details on the timing synchronization are provided in Sec.~\ref{sec2.5}. The spatial overlap of the three lasers was controlled by dichroic mirrors outside the collinear beamline and aligned along beamline’s center axis verified using phosphor screens and iris diaphragms. Since precise spatial overlap between the three laser beams and the pulse ion beam is crucial for achieving high resonance-ionization efficiency, fine adjustments of the optical paths were performed during the measurement to maximize the resonance-ionization ion counts. Moreover, laser power is one of the key factors that affect the measured spectral resolution, primarily due to power broadening. During the experiment, the powers of the three-step lasers were carefully controlled outside the laser cavities using either neutral-density filters or combinations of half-wave plates and polarization beam splitters (PBS).\par

\section{Results of online commissioning}\label{sec3}
\quad The present online commissioning experiment aims to address two central technical issues: whether the RFQ-cb can effectively suppress the large energy spread of the radioactive beam at BRIF, and whether the PLASEN system can realize high-resolution and high-sensitivity laser spectroscopy measurement of unstable nuclei. From high-resolution HFS spectra measured for both stable and unstable Rb isotopes at BRIF, we demonstrate here that the RFQ-cb markedly improves the beam quality and that the PLASEN system enables laser spectroscopy with excellent resolution and sensitivity, as detailed in the following.

\subsection{Evaluation of the ion-beam energy spread at BRIF}\label{sec3.1}

\quad Our earlier studies revealed that the ion beam from BRIF exhibits a relatively large energy spread~\cite{onlineCLS}, which significantly limits the spectral resolution of laser spectroscopy experiments and consequently affects the precision of the extracted nuclear properties. To this end, we performed a direct evaluation of the ion-beam energy spread by monitoring the voltage stability of the BRIF HV platform in this work.

\begin{figure*}[htbp]
    \centering
    \includegraphics[width=0.98\textwidth]{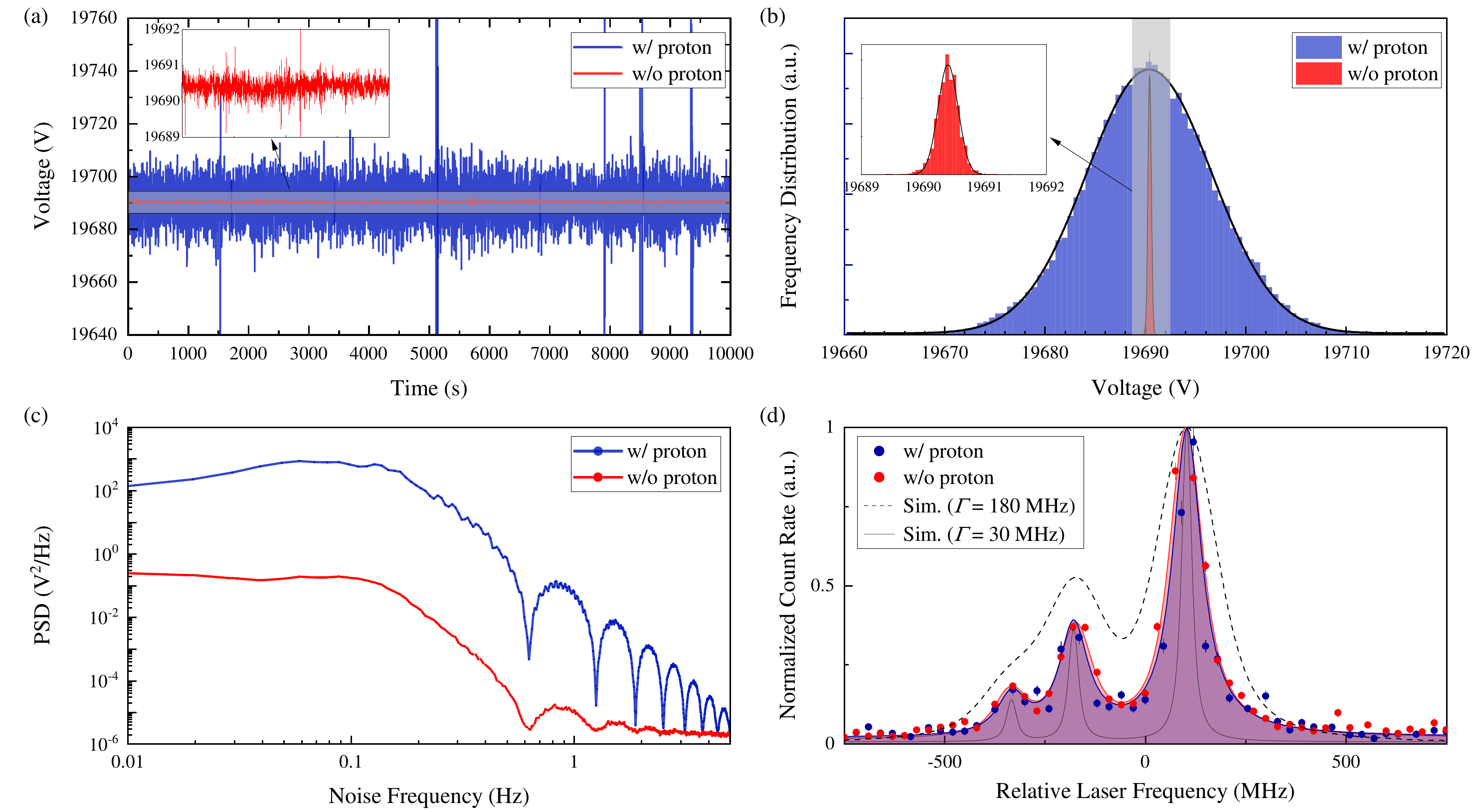}
    \captionsetup{justification=raggedright}
    \caption{(a) Energy fluctuation and (b) voltage distributions of ion-beam delivery platform at BRIF, with proton (blue) and without proton (red) on target. The vertical axis were presented in arbitrary units~(a.u.) for panel (b). (c) Power spectral density (PSD) of noise in the 20-kV HV with proton (blue) and without proton (red) on target, exhibiting typical response characteristics of a moving average filter. Under proton irradiation, the HV noise with proton shows a significant increase at all frequencies. (d) High resolution spectrum of $\mathrm{^{87}Rb}$ measured under the conditions with (blue) and without proton (red) on target. The dashed line represents the simulated spectrum after accounting for Doppler broadening from the $22.04$~eV energy spread, while the solid line shows a simulated spectrum with a best resolution achievable of about 30~MHz.  $\mathit{\Gamma}$ is the total FWHM of the spectral line. See text for details. }
    \label{fig2}
\end{figure*}

\quad During proton irradiation of the target, intense ionizing radiation strongly influences the corona current of the BRIF HV platform. The high-intensity proton beam (up to a few tens of $\upmu$A) is unstable and can induce substantial fluctuations in the corresponding platform voltage $U$, leading to considerable variations in the extracted ion beam energy $E_k = qU$. To semi-quantitatively evaluate the magnitude of the ion-beam energy spread at BRIF, a high-precision voltage divider (Ohm-Labs, KV-30A) was connected to the 20-kV main HV platform of BRIF (see Sec.~\ref{sec2.1}). The platform voltage was continuously monitored and recorded using a digital multimeter (EHENG, EH-ME010) operated at a sampling rate of 10~Hz. To ensure stable and high-resolution measurements, the multimeter's built-in moving average filter was enabled with a filtering strength (i.e., number of averaging points) set to 14, corresponding to a -3~dB cutoff frequency of approximately 0.32~Hz~\cite{Electronics}. Using this configuration, the actual voltage of the 20-kV HV platform at BRIF was measured under conditions both with and without proton irradiation. \par

\quad Figure~\ref{fig2} presents a typical result of the 20-kV platform voltage measurement recorded over a period of about three hours, where blue and red lines correspond to measurements with and without the proton irradiation, respectively. It is evident that the stability of the 20-kV HV deteriorates significantly in the presence of the proton irradiation. To provide a more intuitive and quantitative comparison of the voltage fluctuations, the recorded data were plotted as probability-distribution histograms, as shown in Fig. ~\ref{fig2}~(b) corresponding to the projection of Fig.~\ref{fig2}~(a) onto the vertical axis. As the measured voltage data exhibit a nearly symmetric Gaussian distribution under both conditions, the histograms were fitted with Gaussian profiles. The extracted voltage spreads (FWHM) of the 20-kV HV are 14.69(8)~V (with protons) and 0.384(2)~V (without protons), indicating that proton irradiation significantly enlarges the beam energy spread by approximately a factor of 40. Based on the time-domain data shown in Fig.~\ref{fig2}~(a), the power spectral density (PSD) of the voltage noise was obtained via Fourier analysis, as presented in Fig.~\ref{fig2}~(c). Compared with the case without proton irradiation, the noise level at all frequencies increases markedly under proton irradiation. In addition, since the digital multimeter was configured with built-in filters, the measured noise characteristics exhibit a typical frequency response of a moving-average filter~\cite{Electronics}, with the high-frequency components being largely suppressed.\par
\quad Constrained by the experimental conditions, the voltage fluctuations of the first 10-kV HV platform at BRIF could not be directly monitored. We therefore assume that the voltage fluctuation scales proportionally with the applied voltage. Based on this assumption, the total energy spread of the 30-keV ion beam provided by BRIF is estimated to be at least about 22.04(12)~eV (FWHM). It should be noted that, because the built-in filter of the multimeter strongly suppresses higher-frequency fluctuations above its cutoff frequency, the estimated 22.04-eV energy spread of the ion beam at 30~keV should be regarded as a lower limit.\par

\subsection{RFQ-cb performance for energy-spread reduction of ion beam at BRIF}\label{sec3.2}
\quad By comparing the measured HFS spectra of $^{87}$Rb isotopes with or without proton irradiation of the ISOL target while monitoring the ion-beam energy stability at BRIF (see~\ref{sec3.1}), the capability of the RFQ-cb to reduce the ion-beam energy spread can be semi-quantitatively assessed.\par

\quad Using the formalism in Ref.~\cite{offlineRFQ}
\begin{equation}
  \mathit{\Gamma}_{\rm D} = \frac{\widetilde\nu_0{\delta E}}{\sqrt{2Emc^2}},
\end{equation}
where $\widetilde\nu_0$ corresponds to the $\mathrm{D_2}$ transition frequency of Rb, the Doppler broadening arising from an energy spread of $\delta E=22.04$~eV in a $E=30$~keV ion beam is calculated to be approximately $\mathit{\Gamma}_{\rm D}=121$~MHz. Although significant spectral broadening is expected from the ion-beam energy spread at BRIF under proton irradiation, the RFQ-cb is expected to effectively mitigate the large energy spread, ensuring that the resolution of the measured HFS spectra remains unaffected by proton irradiation. The performance of the RFQ-cb in reducing the beam energy spread was characterized by comparing the measured HFS spectra of the $\mathrm{D_2}$ transition of stable $\mathrm{^{87}Rb}$ under nearly identical experimental conditions (laser powers and timing), but with and without proton irradiation, as shown in Fig.~\ref{fig2}~(d). By fitting the spectra, the FWHM values of the resonance peaks obtained with and without proton irradiation were extracted to be about 86(6)~MHz and 100(6)~MHz respectively, which are essentially comparable. To facilitate a more intuitive visual comparison, a simulated spectrum is also included in Fig.~\ref{fig2}~(d) assuming a spectral resolution of $\mathit{\Gamma}\sim$180~MHz derived from the combined effects of the linewidth measured without proton irradiation and the additional 121~MHz Doppler broadening estimated from $22.04$~eV energy spread.

\quad The slightly better resolution observed under proton irradiation is somewhat counterintuitive, but can be readily explained. Following proton irradiation of the target, a substantial increase in the $\mathrm{^{87}Rb}$ ion beam intensity was observed. To prevent overfilling of the RFQ-cb, the beam intensity was largely reduced from BRIF. As a result, the potential space-charge effect inside the RFQ was alleviated, leading to a slight improvement in the spectral resolution. Besides, it is worth emphasizing that, to ensure sufficient resonance-ionization efficiency, the laser powers used in this test were set to a compromise level, and therefore a certain degree of laser power broadening in the spectral resolution remained. However, as pointed out in our previous work~\cite{offlinePLASEN}, the spectral resolution can be further improved simply by reducing the laser power, though at the cost of resonance-ionization efficiency. A simulated spectrum with an achievable resolution of about 30~MHz FWHM is also shown in Fig.~\ref{fig2}~(d) for reference. For neutron-rich isotopes with very low yields, a balance must thus be considered between pursuing high resolution and maintaining adequate sensitivity. \par

\subsection{High resolution HFS spectra of unstable Rb isotopes}\label{sec3.2}
\begin{figure}[htbp]
    \centering
    \includegraphics[width=0.98\columnwidth]{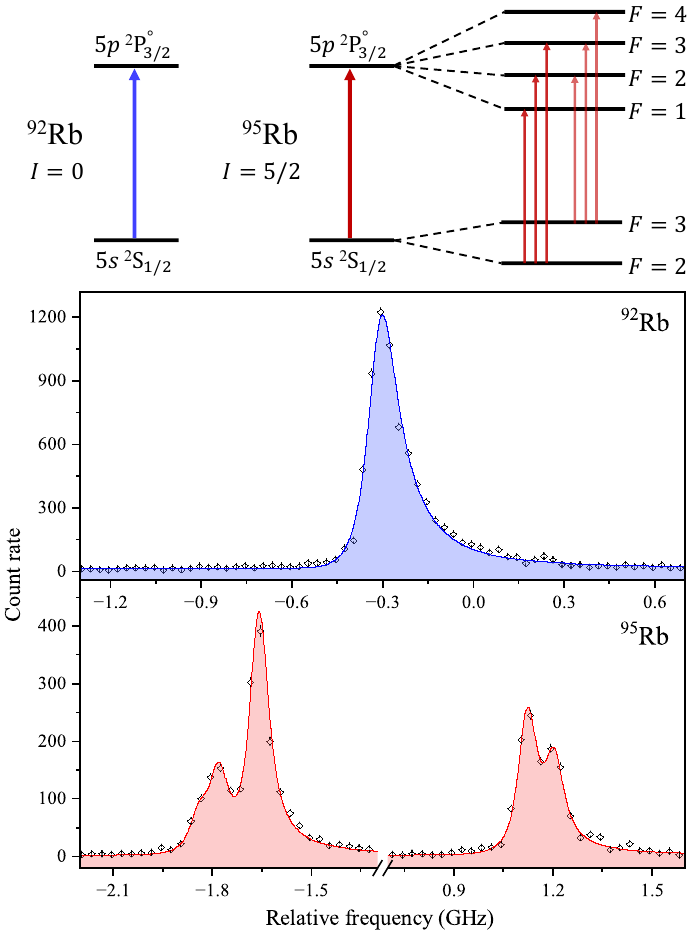}
    \captionsetup{justification=raggedright}
    \caption{Measured high resolution HFS spectra in the unstable $\mathrm{^{92,95}Rb}$ fitted with a asymmetric pseudo-Voigt line profile (solid line). See text for details.}
    \label{fig3}
\end{figure}

\quad Owing to the high-quality bunched ion beam provided by the RFQ-cb, the overall beam transmission through the CLS beamline could reach 80\%, the same level as that achieved with the offline PLASEN system~\cite{offlinePLASEN}. This high-quality bunched beam also enabled us to successfully perform high-resolution and high-efficiency laser spectroscopy measurement of a series of radioactive Rb isotopes using the collinear resonance ionization spectroscopy technique, including the short-lived $\mathrm{^{92}Rb}$ ($T_{1/2}=4.48(3) ~\mathrm{s}$) and $\mathrm{^{95}Rb}$ ($T_{1/2}=377.7(8) ~\mathrm{ms}$). Figure~\ref{fig3} shows the typical HFS spectra obtained for $\mathrm{^{92,95}Rb}$ isotopes. Since $\mathrm{^{95}Rb}$ has a nuclear spin of 5/2, the $\mathrm{D_2}$ transitions give rise to six HFS components due to the coupling between nuclear spin $I$ and electronic total angular momentum $J$. In contrast, the single resonance peak observed for $\mathrm{^{92}Rb}$ reflects its zero nuclear spin. \par
\quad By evaluating the ion counts released from the RFQ-cb and the maximum detected ion signals at the resonance peak of the radioactive $^{92}$Rb isotope, we determined an efficiency of approximately 1:200 achieved at a spectral resolution of about 100~MHz, including contributions from ion beam transmission, CEC neutralization, laser resonance ionization and efficiency of ion detector. This performance matches that obtained with our offline PLASEN system and is comparable to the current state-of-the-art CRIS setup at CERN-ISOLDE~\cite{CRIS}. Moreover, with a transmission efficiency exceeding 30\% from mass separation to FC2 (after the RFQ-cb), as discussed in Sec.\ref{sec2.2}, an overall efficiency of about 1:666 was achieved from ion counts after mass separation to the detected resonance at a spectral resolution of 100~MHz. This performance enables laser spectroscopy measurement of more neutron-rich Rb isotopes with production yields as low as $~\sim$100~pps within reasonable measurement times. \par

\quad From Fig.~\ref{fig3}, we also note that the observed resonance peaks exhibit asymmetry, which is commonly known to be attributed to two main factors. The first is the energy loss due to the population of higher-lying states during the charge-exchange process or through additional inelastic collisions~\cite{CEC-Asymm}. The second is AC Stark shifts induced by the high-power second- and third-step laser pulses, which may arrive close in time to the first-step pulse~\cite{ACStarkPRA,ACStark}. However, investigations performed with the offline PLASEN system~\cite{offlinePLASEN}, where the CEC heating temperature was significantly reduced and the timing of the second- and third-step laser pulses was delayed, ruled out both factors as the dominant source of the observed asymmetry. Instead, we found that this asymmetry is strongly correlated with the density of the helium buffer gas introduced into the RFQ-cb, which is located directly upstream of the CLS beamline. To fully clarify the underlying mechanisms linking the buffer gas density to the asymmetric resonance profile, further systematic investigations are still required.

\quad Under the present conditions, we adopted an asymmetric pseudo-Voigt profile~\cite{SATLAS2} to account for the observed asymmetry, for fitting the HFS spectra of $\mathrm{^{87,92,95}Rb}$. The extracted atomic parameters, including the magnetic-dipole and electric-quadrupole HFS constants ($A$ and $B$) for the $5s~^2\mathrm{S}_{1/2}$ and $5p~^2\mathrm{P}_{3/2}^{\circ}$ states of $\mathrm{^{87,95}Rb}$, as well as the isotope shifts of $\mathrm{^{87,95}Rb}$ relative to the reference isotope of $\mathrm{^{92}Rb}$, are summarized in Table~\ref{tab1}. Despite the asymmetric resonance profiles, the extracted physical parameters show excellent agreement with literature values~\cite{thibault1981hyperfine} (Table~\ref{tab1}), further confirming the performance and reliability of the PLASEN system.\par

\begin{table*}[t!]
\captionsetup{justification=raggedright}
\caption{\footnotesize{HFS constants for $^{87,95}$Rb and isotope shifts of $^{87,95}$Rb relative to the reference isotope of $^{92}$Rb, compared to the literature values~\cite{thibault1981hyperfine}. All values are given in the unit of MHz.}}\label{HFS}
\setlength{\tabcolsep}{7pt}
\setlength{\extrarowheight}{4pt}
\centering
\begin{tabular}{cc|cccc|cccc}
\toprule
& &  \multicolumn{4}{c|}{\mbox{This work}}  & \multicolumn{4}{c}{\mbox{Literature values~\cite{thibault1981hyperfine}}}\\
\hline
A &$I^{\pi}$& $A(\mathrm{S}_{1/2})$& $A({P}_{3/2}^{\circ})$&$B({P}_{3/2}^{\circ})$&$ \rm{IS} $& 
$A(\mathrm{S}_{1/2})$&$A({P}_{3/2}^{\circ})$ &$B({P}_{3/2}^{\circ})$&$ \rm{IS}$\\
\hline
87 & $3/2^{-}$ & 3414.9(3) & 84.97(27) & 11.2(7) & 317.9(8) & 3415.9(20) & 84.29(50)  & 12.2(20) & 320.4(52) \\
95 & $5/2^{-}$ & 994.1(9) & 25.0(6)  & 18.4(19) & $-177.4(29)$ &  993.7(25) & 24.0(9)  & 19.8(61)  & $-175.5(65)$  \\
\bottomrule
 \label{tab1}
\end{tabular}
\end{table*}

\section{Summary and Outlook}

\quad In this work, we have successfully installed and commissioned the complete PLASEN system at BRIF, consisting of a RFQ-cb and a collinear resonance ionization spectroscopy setup. Using both stable and radioactive Rb beams, we have demonstrated that the RFQ-cb effectively suppresses the large beam-energy spread of ion beams delivered by BRIF, enabling the delivery of high-quality, bunched radioactive ion beams for precision laser spectroscopy. Under these optimized conditions, high-resolution (100~MHz FWHM) and high-sensitivity (1:200) laser spectroscopy of unstable nuclei has been achieved. These results validate the capability of PLASEN for nuclear-structure studies of neutron-rich isotopes with very low production yields and establish it as a powerful platform for high-resolution and high-sensitivity laser spectroscopy at RIB facilities. Moreover, building upon this successful commissioning, the PLASEN system may provide broad opportunities that bridge nuclear, atomic, and precision-measurement sciences. \par

\quad In the domain of RIB physics, BRIF is capable of delivering a large range of radioactive ion beams in the medium-mass region produced via nuclear fission~\cite{NuclPhysatBRIF}. Many of these nuclei are expected to exhibit exotic structural phenomena, and some lie close to the predicted $r$-process path responsible for synthesizing roughly half of the nuclei heavier than iron~\cite{PhysofExoticNuclei}. With the demonstrated performance of PLASEN, nuclear spins, magnetic dipole moments, electric quadrupole moments, and charge radii can be measured for these isotopes. Experiments on neutron-rich $\mathrm{^{99,100}Rb}$ and their possible long-lived isomers are underway, aiming to investigate the deformation around $Z=40$ and $N=60$~\cite{Rb98}. 
 Additionally, the online PLASEN system enables laser-assisted decay spectroscopy with purified isomeric beams. For neutron-rich nuclei in the fission-fragment region, numerous long-lived isomers exist. These isomers are of great importance for both nuclear-structure research and nuclear-data evaluation. High-resolution collinear resonance ionization laser spectroscopy can selectively resolve and separate long-lived nuclear isomeric states due to the presence of isomer shifts, thereby providing high-purity isomeric beams for decay spectroscopy experiments~\cite{CRISDSSFr1,CRISDSSFr2,CRISDSS}. A dedicated decay station has recently been installed downstream of the PLASEN system at BRIF to study isomeric decays in the neutron-rich medium-mass region. \par

\quad The PLASEN system also opens new opportunities at BRIF for spectroscopic studies of atoms and molecules containing unstable nuclei, contributing to research on fundamental symmetries. In atomic and molecular systems, weak interactions between nuclei and electrons can induce parity non-conservation (PNC), which provide sensitive probes of nuclear electroweak properties~\cite{AMONewPhysics}. While the most precise atomic PNC measurement has been achieved in $\mathrm{^{133}Cs}$~\cite{CsAPV}, larger effects are predicted in certain unstable species (e.g., $\mathrm{^{221}Fr}$~\cite{FrPNC}), though some key atomic spectroscopic data remain scarce. With access to short-lived isotopes at BRIF, the PLASEN system offers a promising platform for spectroscopic investigations on those candidates. Besides, searches for permanent electric dipole moments (EDMs), closely related to CP violation and the matter–antimatter asymmetry~\cite{EDMReivew1,EDMReview2}, represent another frontier in research on fundamental symmetries. Radioactive molecules containg heavy octupole-deformed nuclei (e.g., $\mathrm{^{225}Ra}$) are predicted to offer enhanced sensitivity to nuclear EDMs ~\cite{RadioactiveMolecules}, but most of their spectroscopic structures remain largely unexplored. Collinear resonance ionization spectroscopy of RaF~\cite{RaFNat,RaFPRL,RaFPRA,RaFNatCommu,fvsy-v1q6,doi:10.1126/science.adm7717} and AcF~\cite{AcFCRIS} has been demonstrated at CERN–ISOLDE. In the PLASEN system, the RFQ-cb enables in-trap formation of radioactive molecules, including polyatomic species. Following successful production and laser spectroscopy studies of BaF in our offline system~\cite{Molecular}, online investigations of RaF and RaOH are planned, paving the way for future precision tests of fundamental symmetries with radioactive molecules.\par

\begin{acknowledgments}
This work was supported by the National Key R\&D Program of China (2023YFE0101600, 2023YFA1606403, 2022YFA1605100, 2023YFA1607001), the National Natural Science Foundation of China (12027809, 12350007, and 12305122), and New Cornerstone Science Foundation through the XPLORER PRIZE. The authors would like to thank the members of the CRIS and COLLAPS collaboration for their valuable discussions and support in the development of the PLASEN system.
\end{acknowledgments}

\bibliographystyle{apsrev4-2}
\bibliography{Reference}

\end{document}